\documentclass[showpacs,twocolumn,aps]{revtex4}
\usepackage{amssymb}
\usepackage{amsmath}
\usepackage{graphicx}
\usepackage{lscape}
\usepackage{booktabs}
\begin{document}
\title{Light anti-nuclei production in pp collisions at $\sqrt{s}$=7 and 14
TeV}
\author{Yu-Liang Yan$^{1}$, Gang Chen$^{2}$, Xiao-Mei Li$^{1}$,
Dai-Mei Zhou$^{3}$, Mei-Juan Wang $^{2}$, Shou-Yang Hu$^{1}$, Li Ye$^{1}$,
Ben-Hao Sa$^{1,3}$}
\address{
1 China Institute of Atomic Energy, P.O. Box 275(18), Beijing 102413, China\\
2 Physics Department, China University of Geoscience, Wuhan 430074, China\\
3 Institute of Particle Physics, Huazhong Normal University, Wuhan 430082,
  China}

\begin{abstract}
A dynamically constrained coalescence model based on the phase space
quantization and classical limit method was proposed to investigate
the production of light nuclei (anti-nuclei) in non-single
diffractive (NSD) pp collisions at $\sqrt{s}$=7 and 14 TeV. This
calculation was based on the final hadronic state in the PYTHIA and
PACIAE model simulations, the event sample consisted of 1.2$\times
10^8$ events in both simulations. The PACIAE model calculated
$\overline D$ yield of 6.247$\times 10^{-5}$ in NSD pp collisions at
$\sqrt{s}$=7 TeV is well comparing with the ALICE rough datum of
5.456$\times 10^{-5}$. It indicated the reliability of proposed
method in some extent. The yield, transverse momentum distribution,
and rapidity distribution of the $\overline D$, $^3{\overline{He}}$,
and $_{\overline\Lambda} ^3{\overline H}$ in NSD pp collisions at $\sqrt{s} $=7 and 14 TeV
were predicted by PACIAE and PYTHIA model simulations. The yield
resulted from PACIAE model simulations is larger than the one from
PYTHIA model. This might reflect the role played by the
parton and hadron rescatterings.

\end{abstract}

\maketitle

\section{Introduction}
The nucleus-nucleus collisions at top RHIC energy produced an initial hot
and dense matter (quark-gluon mater, QGM) and has been interpreted
as a strongly coupled quark-gluon plasma (sQGP) \cite{brah,phob,star,phen}.
This is nearly a perfect liquid composed of quarks, antiquarks, and gluons
bur is not a free gas-like quark-gluon plasma (fQGP) expected by theorists
and experimentalists long time ago.

We have proposed physical conjectures to theoretically explore the property
differences among the QGM formed in pp and/or nucleus-nucleus collisions
at ultra-relativistic energies \cite{sa3}. These conjectures are: (1) The
fraction of gluon in yield and momentum increases with increasing reaction
energy. (2) The quark flavors ($u, d, s, ...$) approach equilibrium with
increasing reaction energy. (3) The antiparton to parton ($u, d, s, ...$
quarks) ratio increases with increasing reaction energy. In short, the
population of $u, d, s, c, ...$ quarks and the antiquark to quark ratio
approach equilibrium (balance) and unity with increasing
reaction energy, respectively.

Recently STAR reported the observation of ``an equilibrium in coordinate and
momentum space populations of up, down, and strange quarks and antiquarks" by
``The measured yields of $^3_{\Lambda} H$ ($^3_{\bar\Lambda}\overline H$) and
$^3He$ ($^3\overline{He}$) are similar" or by the observation of $^3_{\bar
\Lambda}\overline H/^3_{\Lambda} H$ is close to $^3\overline{He}/^3He$ in Au+Au
collisions at top RHIC energy \cite{star2}. Whether this STAR observation
means the initial fireball created in this collision is really a fQGP may
be debated.

The investigation of anti-nuclei has great meaning in the nuclear and
particle physics, the astrophysics, and even in the cosmology. Recently ALICE
published their preliminary results of $\overline D$ production in pp
collisions at $\sqrt s$= 7 TeV \cite{alice} at nearly ten months later than
STAR publishing their measurements for $_{\overline\Lambda} ^3{\overline H}$ production in Au+Au
collisions at $\sqrt {s_{NN}}$=200 GeV \cite{star2}. However, because of quite
low multiplicity, the study of anti-nuclei production is very hard both in
experiment and theory.

So far the report about the formation of QGM in early stage of pp collisions
at RHIC energy is still absent in our knowledge. However, there were studies
about the measurable of flow parameters and the elliptic flow signature of
the QGP phase transition in high multiplicity (energy) pp collisions
\cite{casa,huma,boze,orto,chau,zhou}. So one could not rule out the QGM
formation possibility in pp collisions at LHC or higher energies \cite{arme}.

In this paper the PYTHIA model \cite{sjo2} and PACIAE model \cite{sa2} were
used to calculate the final hadronic state in non-single diffractive (NSD)
pp collisions at $\sqrt s$=7 and 14 TeV. That was followed by the
calculations for the production of light anti-nuclei in the dynamically
constrained coalescence model based on the phase space quantization and
classical limit method. The PACIAE model calculated $\overline D$ yield of
6.247$\times 10^{-5}$ in NSD pp collisions at $\sqrt{s}$=7 TeV is well
comparing with the ALICE rough datum of 5.456$\times 10^{-5}$. This may
indicate the reliability of the proposed method in some extent. The parton
and hadron rescattering effects were analyzed by comparing the PACIAE results
with the PYTHIA one. It turned out that their role is un-negligible.

\section {MODELS}
PYTHIA is a model for high energy hadron-hadron ($hh$) collisions \cite{sjo2}.
The parton and hadron cascade model, PACIAE \cite{sa2}, is based on PYTHIA.
In the PYTHIA model a hh collision is decomposed into parton-parton
collisions. The hard parton-parton collision is described by lowest leading
order perturbative QCD (LO-pQCD) parton-parton interactions with the
modification of parton distribution function in a hadron. The soft
parton-parton collision, non-perturbative phenomenon, is considered
empirically. The initial- and final-state QCD radiations and multiparton
interactions are considered. So the consequence of a $hh$ collision is a
partonic multijet state composed of di-quarks (anti-diquarks), quarks
(antiquarks), and gluons, besides a few hadronic remnants. It is then
followed by the string construction and fragmentation, one obtains a hadronic
final state for a $hh$ (pp) collision eventually.

For the pp collisions the PACIAE model is different from the PYTHIA in
follows:
\begin{enumerate}
\item The string fragmentation is switched-off and the di-quarks
(anti-diquarks) are broken randomly into quarks (antiquarks). So the
consequence of a pp collision is a initial state of quarks, antiquarks, and
gluons, besides a few hadronic remnants. This partonic initial state is
regarded as the hot QCD matter (QGM) formed in the relativistic pp
collisions.

\item The parton rescattering is introduced. In this stage the rescattering
among partons in QGM is considered by the 2 $\rightarrow$ 2 LO-pQCD
parton-parton interaction cross sections \cite{comb}. However, a $K$ factor
is introduced to consider the higher order and non-perturbative corrections.
The effective strong coupling constant is assumed to be $\alpha_s$=0.47.
A parton colour screen mass $\mu$=0.63 GeV is introduced to avoid the
divergence. Integrate the differential cross sections above, the total cross
section of the parton collision is obtained. Then the parton rescattering
is simulated by the Monte Carlo method.

\item The hadronization is proceeded after parton rescattering. The partonic
matter can be hadronized by the Lund string fragmentation regime \cite{sjo2}
and/or phenomenological coalescence model \cite{sa2}.
\item At last the hadron rescattering is added. In this stage the hadronic
matter after hadronization suffers rescattering. It is dealt with by the
usual two-body collision method \cite{sa1}, until the hadronic freeze-out
(the hh collision pair is exhausted).
\end{enumerate}
In short, the PACIAE model consists of the parton initialization, parton
evolution (rescattering), hadronization, and hadron evolution (rescattering)
four stages.

STAR \cite{star2} has found in Au+Au collisions at top RHIC energy that
``The measured $_{\bar\Lambda}^3\overline{H}/_\Lambda ^3H$ and
$^3\overline{He}/^3{He}$ ratios are consistent with the interpretation that
the $_{\bar\Lambda}^3\overline{H}$ and $_\Lambda ^3H$ are formed by
coalescence of ($\overline\Lambda+\overline p+\overline n$) and ($\Lambda+p
+n$), respectively." Obviously, this simplest coalescence assumption is
utilized only for ratio and not for yield.

In the theoretical studies, the yield of light nuclei (anti-nuclei) was
always calculated in two steps: In the first step the nucleons and hyperons
were calculated by transport model. Then the light nuclei (anti-nuclei) were
calculated by the analytical coalescence model \cite{grei,chen,ma} and/or
the statistical model \cite{pop}. This meant the formation of nuclei
(hypernuclei) was not dynamically treated continuously but as a final state
interaction separately. The above analytical coalescence model relies upon
the Wigner function \cite{wign} constructed according to the assumed wave
function of light nuclei (anti-nuclei) \cite{ma}. The statistical
model relies upon the equilibrium and temperature assumptions as well as the
restriction on projectile-like fragment in the intermediate energy heavy ion
collisions. In this paper we proposed a dynamically constrained coalescence
model based on the phase space quantization and classical limit method to
calculate the yield of light nuclei (anti-nuclei) after the transport model
simulation directly.

One knew in statistical mechanics \cite{kubo} that, for a system with three
dimensions in the state definition the set of microscopic states held in a
volume element $\Delta\Gamma$ corresponds, in the limit of $h\rightarrow 0$,
to a set of
\begin{equation}
\frac{\Delta\Gamma}{h^3}
\label{phas}
\end{equation}
quantum states. Of course, this correspondence is an approximation as long as
$h$ remains finite. In the Eq.~\ref{phas} $\Delta\Gamma$ reads
\begin{equation}
\Delta\Gamma\equiv\Delta\vec q\Delta\vec p
\end{equation}
where $\vec q$ and $\vec p$ stand for the three position and three momentum, 
respectively. In other word, a $h^3$ volume element in the three
dimensions phase space corresponds to a state of the system.

\section{Calculations and results}
Taking the system of $_{\overline\Lambda} ^3{\overline H}$ as an example, a configuration
consists of $\bar p$, $\bar n$, and $\bar\Lambda$ in a single event of the
final hadronic state from transport model simulation could be expressed as
\begin{equation}
C_{\bar p\bar n\bar\Lambda}(q_1,q_2,q_3;\vec p_1,\vec p_2,\vec p_3),
\label{conf}
\end{equation}
where subscripts 1, 2, and 3 were a shorthand for $\bar p$, $\bar n$, and
$\bar\Lambda$, respectively and $q_1$ refers to the distance between $\bar p$
and the center-of-mass of $\bar p$, $\bar n$, and $\bar\Lambda$, for
instance. This configuration contributed an partial yield of
\begin{equation}
\delta_{123}=\left\{
  \begin{array}{ll}
  1 \hspace{0.4cm} if \hspace{0.2cm} m_{inv}\leq m_0\pm\Delta m, \hspace{0.2cm}
    q_1\leq R_0, \\
    \hspace{1.2cm} q_2\leq R_0, \hspace{0.2cm} q_3\leq R_0; \\
  0 \hspace{0.4cm}otherwise;
  \end{array}
  \right.
\label{yield5}
\end{equation}
to the $_{\overline\Lambda} ^3{\overline H}$. In the above equation
\begin{equation}
m_{inv}=[(E_1+E_2+E_3)^2-(\vec p_1+\vec p_2+\vec p_3)^2]^{1/2}
\label{yield3}
\end{equation}
is the invariant mass, $m_0$ and $R_0$ stand for the rest mass and radius of
the $_{\overline\Lambda} ^3{\overline H}$, $\Delta m$ refers to the allowed mass uncertainty. The
total yield of $_{\overline\Lambda} ^3{\overline H}$ in a single event is the sum of above partial
yield first over the Eq. \ref{conf} type configurations and then over the
configuration types obtained by the combination among subscripts 1, 2, and 3.
An average over events is required at last.

\begin{table}[htbp]
\caption{Particle yield in NSD pp collisions at $\sqrt{s}$=0.2 TeV.}
\begin{tabular}{cccc}
\hline
    & $\rm{STAR}^a$ &PACIAE & PYTHIA\\
\hline
$k^+$ &0.140$\pm$0.010$$ &0.137 &0.125\\
$k^-$ &0.137$\pm$0.009 &0.122 &0.115\\
$\Lambda$ &0.0385$\pm$0.0036 &0.0382 &0.0309\\
$\overline{\Lambda}$ &0.0351$\pm$0.0033 &0.0381 &0.0312 \\
\hline
\multicolumn{4}{l}{$^a$ The STAR data were taken from \cite{star3}}\\
\end{tabular}
\label{paci1}
\end{table}

\begin{table}[htbp]
\caption{Hadron and light nuclei (anti-nuclei) yields in
NSD pp collisions at $\sqrt s$=7 and 14 TeV calculated by final hadronic
state in the PACIAE and PYTHIA model simulations.}
\begin{tabular}{ccccc}
\hline
   &\multicolumn{2}{c}{PACIAE} &\multicolumn{2}{c}{PYTHIA}\\
\cmidrule[0.25pt](l{0.03cm}r{0.03cm}){2-3}
\cmidrule[0.25pt](l{0.03cm}r{0.03cm}){4-5}
&7 TeV &14 TeV &7 TeV &14 TeV\\
\hline
$k^+$ &4.563 &5.576 &3.802 &4.946 \\
$k^-$ &4.416 &5.331 &3.689 &4.778\\
$p$ & 4.152 &4.678 &3.491 &4.074\\
$\overline p$ &3.040 &3.588 &2.472 &3.078\\
$n$ &4.094 &4.677 &3.397 &4.107\\
$\overline n$ &3.336 &3.938 &2.565 &3.335\\
$\Lambda$ &1.648 &1.940 &1.285 &1.608\\
$\overline{\Lambda}$ &1.518 &1.769 & 1.136&1.386\\
$ D^{a}$ & 6.906E-05  & 9.111E-05  &5.586E-05   &6.724E-05  \\
$\overline D^{a}$ & 6.247E-05&8.553E-05 &4.852E-05 &6.048E-05\\
              & 5.456E-05$^{b}$& & & \\
$_{\Lambda} ^3H^{c}$ &2.547E-07 &3.814E-07   & 1.833E-07  &9.677E-08  \\
$_{\overline\Lambda} ^3{\overline H}^{c}$ &2.453E-07 &3.305E-07&1.000E-07 &1.048E-07  \\
$^3{He}^{c}$ &2.453E-07   &3.898E-07   & 1.500E-07  & 1.774E-07  \\
$^3{\overline{He}}^{c}$ &2.642E-07 &4.407E-07   & 1.417E-07 &2.419E-07  \\
\hline
\multicolumn{5}{l}{$^a$ Calculated with $\Delta m$=0.0005 GeV.} \\
\multicolumn{5}{l}{$^b$ Estimated from Fig. 4 in \cite{alice}.}\\
\multicolumn{5}{l}{$^c$ Calculated with $\Delta m$=0.005 GeV.}
\end{tabular}
\label{paci}
\end{table}

\begin{figure*}[htb!]
\includegraphics[width=0.8\textwidth]{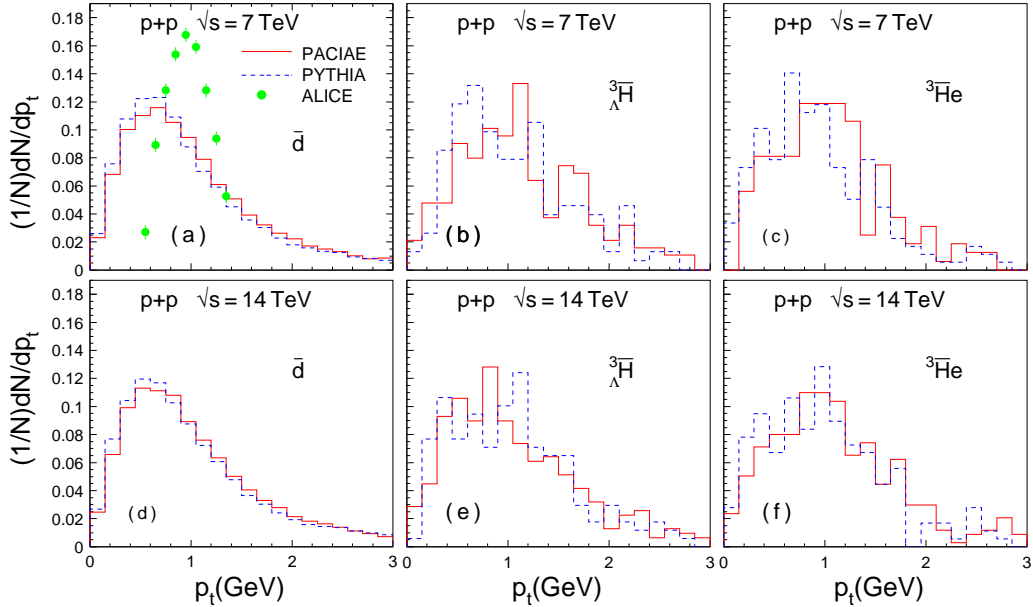}
\caption{(Color online) Light anti-nuclei transverse momentum distributions
in the NSD pp collisions at $\sqrt s$=7 and 14 TeV. The panels
(a), (b), and (c) were for $\overline D$ ($\Delta m$=0.0005 GeV),
$_{\overline\Lambda} ^3{\overline H}$ ($\Delta m$=0.005 GeV), and $^3{\overline{He}}$ ($\Delta m$=0.005 GeV),
respectively, in the NSD pp collisions at $\sqrt s$=7 TeV, the panels (d),
(e), and (f) at $\sqrt s$=14 TeV. The red solid and blue dashed histograms were
calculated by final hadronic state in PACIAE and PYTHIA model
simulations, respectively.}
\label{tran}
\end{figure*}

\vspace{2cm}
\begin{figure*}[htb!]
\includegraphics[width=0.8\textwidth]{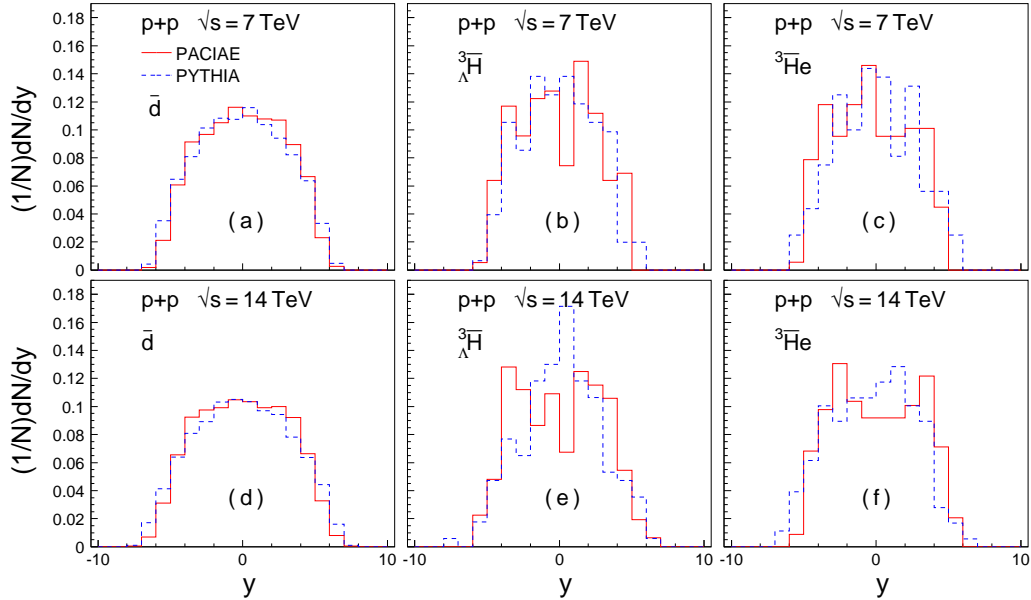}
\caption{(Color online) The same as Fig.~\ref{tran} but for rapidity.}
\label{rapi}
\end{figure*}

The above summation was easily practised by following algorithm: One devised a
three level loops ($i$, $j$, and $k$) over particle list in a single event of
the final hadronic state from transport model simulation. If the particle
code \cite{pdg} (KF code in PYTHIA model \cite{sjo2}) sum of the three
particles ($i$, $j$, and $k$) was equal to -7446 (=-2212-2112-3122) and the
condition of Eq. \ref{yield5} was satisfied then the $_{\overline\Lambda} ^3{\overline H}$ yield
increased by one unit. The loops was then proceeded by jumping to $i$ loop,
the regular loop proceeding otherwise.

In the PYTHIA and PACIAE simulations we assumed that the hyperons heavier
than $\Lambda$ will decay. The model parameters were fixed on the default
values given in PYTHIA model, except the $K$ factor and the parameters of
parj(1), parj(2), and parj(3) (the later three were concerning in hyperon
production \cite{sjo2}) were roughly fitted to the STAR data of $k^+, k^-,
\Lambda$, and $\overline \Lambda$ in NSD pp collisions at $\sqrt{s}$= 0.2 TeV
\cite{star3} as shown in Tab.~\ref{paci1}. The fitted parameter values of 3
(default value is 1 or 1.5), 0.15 (0.1), 0.38 (0.3), and 0.45 (0.4) were used
to calculate the yield of $\overline D$, $^3{\overline{He}}$, and
$_{\overline\Lambda} ^3{\overline H}$ in NSD pp collisions at $\sqrt{s}$=7 and 14 TeV by the final
haronic state in PACIAE and PYTHIA model simulations as shown in Tab.
~\ref{paci}.

One sees in the Tab.~\ref{paci} that:
\begin{itemize}
\item The $\overline D$ yield of 6.247$\times 10^{-5}$ in NSD pp collisions at
$\sqrt{s}$=7 TeV in the PACIAE simulations seems reasonable comparing with
the ALICE rough datum of $\sim 5.456\times 10^{-5}$ (roughly estimated from
Fig. 4 in \cite{alice} where integrating over $p_t$ and dividing by 350 M
triggered events measured). The value in the PACIAE model larger than ALICE
datum may attribute to the full rapidity phase space in the former (the same
in all calculations in both PACIAE and PYTHIA models) bur $|\eta|<0.9$ in the
later.
\item The PACIAE yield of anti-hadron increases with increasing reaction
energy from 7 TeV to 14 TeV in a percentage of $\sim$ 20\%. It is less than 
the light anti-nuclei ($\sim$ 37\% for $\overline D$, $\sim$ 50\% for
$_{\overline\Lambda} ^3{\overline H}$, and $\sim$ 70\% for $^3{\overline{He}}$). This might attribute to
the available phase space increases with increasing reaction energy is stronger 
in the light anti-nuclei production rather than in the anti-hadron production.
\item The yield resulted from PACIAE model simulation is larger than the one
from PYTHIA model simulation at the same reaction energy. This reflects the role
played by the parton and hadron rescatterings.
\end{itemize}

Table~\ref{paci2} showed the feature of average transverse momentum of the
produced light nuclei (anti-nuclei) is not sensitive to the special piece of
nuclei (anti-nuclei) and the reaction energy, like the case of hadron
production. However, the $\langle p_t \rangle$ seems to be $\sim$ 1 Gev/c in light nuclei
(anti-nuclei) production instead of $\sim$ 0.5 Gev/c in hadron production.
This might mean the light nuclei (anti-nuclei) are generated more
isotropically in the momentum phase space.

Figure~\ref{tran} gave the predicted $\overline D$ ($\Delta m$=0.0005 GeV),
$_{\overline\Lambda} ^3{\overline H}$ ($\Delta m$=0.005 GeV), and $^3{\overline{He}}$ ($\Delta m$=0.005 GeV)
transverse momentum distributions in NSD pp collisions at $\sqrt s$= 7 and
14 TeV. The panal $(a)$ was for $\overline D$ where the ALICE data
\cite{alice} were also given by the green circles. One sees in this panal
that the $dN/dp_t$ shape in PACIAE and/or PYTHIA model simulations
is similar to the one in general hadron production. The experimental
Gaussian-like $dN/dp_t$ may be modified after ``efficiency and annihilation
corrections" \cite{alice}. The Fig.~\ref{tran} show that the 1.2$\times10^8$
events are enough for $dN/dp_t$ distribution of $\overline D$ but not enough for
$_{\overline\Lambda} ^3{\overline H}$ and $^3{\overline{He}}$. This was also the reason using
different $\Delta m$ value in the calculations for $\overline D$ and for
$_{\overline\Lambda} ^3{\overline H}$ and $^3{\overline{He}}$. The larger fluctuation are shown in
panels $(b), (c), (e)$, and $(f)$. Globally speaking, the PACIAE $dN/dp_t$
distribution is not so much different form the PYTHIA one.

We also predicted the rapidity distribution for the $\overline D$,
$^3{\overline{He}}$, and $_{\overline\Lambda} ^3{\overline H}$ in NSD pp collisions at $\sqrt s$=7
and 14 TeV in Fig.~\ref{rapi}. In this figure one sees that the global
features shown in $dN/dy$ distribution are quite similar to the $dN/dp_t$
distribution in Fig.~\ref{tran} so need not to repeat.

\begin{table}[htbp]
\caption{Light nuclei (anti-nuclei) average transverse momentum in
NSD pp collisions at $\sqrt s$=7 and 14 TeV calculated by final hadronic
state in the PACIAE and PYTHIA model simulations.}
\begin{tabular}{ccccc}
\hline
    &\multicolumn{2}{c}{PACIAE} &\multicolumn{2}{c}{PYTHIA}\\
\cmidrule[0.25pt](l{0.03cm}r{0.03cm}){2-3}
\cmidrule[0.25pt](l{0.03cm}r{0.03cm}){4-5}
&7 TeV &14 TeV &7 TeV &14 TeV\\
\hline
$ D$ & 0.988& 0.993& 0.941 &0.952 \\
$\overline D$ & 0.999& 1.01&0.955 &0.981 \\
&0.962$^a$ &&0.955$^a$ &\\
$_{\Lambda} ^3H$ &1.07 &1.05 &0.909 &1.01 \\
$_{\overline\Lambda} ^3{\overline H}$ &1.14 &1.05 &1.09 &1.00 \\
$^3{He}$ &1.05 &1.07 &0.983 &1.04 \\
$^3{\overline{He}}$ &0.939 &1.10 &1.05 &1.05 \\
\hline
\multicolumn{5}{l}{$^a$ Estimated according to the Fig. 4 in \cite{alice}.}
\end{tabular}
\label{paci2}
\end{table}

\section{Conclusion}
In summary, the PYTHIA model and the parton and hadron cascade model
PACIAE based on PYTHIA were employed to investigate the production
of light nuclei (anti-nuclei) in NSD pp collisions at $\sqrt s$=7
and 14 TeV. We proposed a dynamically constrained coalescence model
based on the phase space quantization and classical limit method to
calculate the light nuclei (anti-nuclei) by the final hadronic state
in PACIAE (PYTHIA) model simulations. The calculated $\overline D$
yield of 6.247$\times 10^{-5}$ in NSD pp collisions at $\sqrt s$=7
TeV by PACIAE model final hadronic state is quite close to the ALICE
rough datum of 5.456$\times 10^{-5}$ \cite{alice}. It turned out
the reliability of proposed method in some extent. The light nuclei
(anti-nuclei) yield, rapidity distribution, and transverse momentum
distribution in NSD pp collisions at $\sqrt{s}$=7 and 14 TeV were
also predicted by PACIAE and PYTHIA model simulations. The yield resulted
from the PACIAE model simulations is larger than the one from PYTHIA model
which might reflect the role played by the parton and
hadron rescatterings.

\begin{center} {ACKNOWLEDGMENT} \end{center}
Finally, we acknowledge the financial support from NSFC (11075217, 11047142,
and 10975062) in China.

\end{document}